%% file: main.tex
\documentclass[sigconf,screen]{acmart}

\usepackage[utf8]{inputenc}
\usepackage{xspace}
\usepackage{listings}
\usepackage{url}
\usepackage{adjustbox}
\usepackage{hyperref}
\usepackage{float}
\usepackage{cleveref}
\usepackage[htt]{hyphenat}
\usepackage{microtype}

\setcopyright{acmcopyright}
\acmPrice{15.00}
\acmDOI{10.1145/3368089.3417929}
\acmYear{2020}
\copyrightyear{2020}
\acmSubmissionID{fse20demo-p19-p}
\acmISBN{978-1-4503-7043-1/20/11}
\acmConference[ESEC/FSE '20]{Proceedings of the 28th ACM Joint European Software Engineering Conference and Symposium on the Foundations of Software Engineering}{November 8--13, 2020}{Virtual Event, USA}
\acmBooktitle{Proceedings of the 28th ACM Joint European Software Engineering Conference and Symposium on the Foundations of Software Engineering (ESEC/FSE '20), November 8--13, 2020, Virtual Event, USA}

\begin{document}
\input{misc/commands}

\title{\prf: A Framework for Building Automatic Program Repair Prototypes for JVM-Based Languages}

\author{Ali Ghanbari}
\affiliation{
  \institution{University of Texas at Dallas}
  \city{Richardson}
  \state{TX 75080}
  \country{USA}
}
\email{ali.ghanbari@utdallas.edu}

\author{Andrian Marcus}
\affiliation{
  \institution{University of Texas at Dallas}
  \city{Richardson}
  \state{TX 75080}
  \country{USA}
}
\email{amarcus@utdallas.edu}

\begin{abstract}
\prf is a Java-based framework that allows researchers to build prototypes of test-based generate-and-validate automatic program repair techniques for JVM languages by simply extending it with their patch generation plugins.
The framework also provides other useful components for constructing automatic program repair tools, \textit{e.g.,} a fault localization component that provides spectrum-based fault localization information at different levels of granularity, a configurable and safe patch validation component that is 11+X faster than vanilla testing, and a customizable post-processing component to generate fix reports.\\
A demo video of \prf is available at \url{https://bit.ly/3ehduSS}.
\end{abstract}

\ccsdesc[500]{Software and its engineering~Software testing and debugging}

\keywords{Automatic Program Repair, Framework, Fault Localization, Patch Validation}

\maketitle

\begin{figure*}[t!]
    \centering
    \includegraphics[scale=0.4]{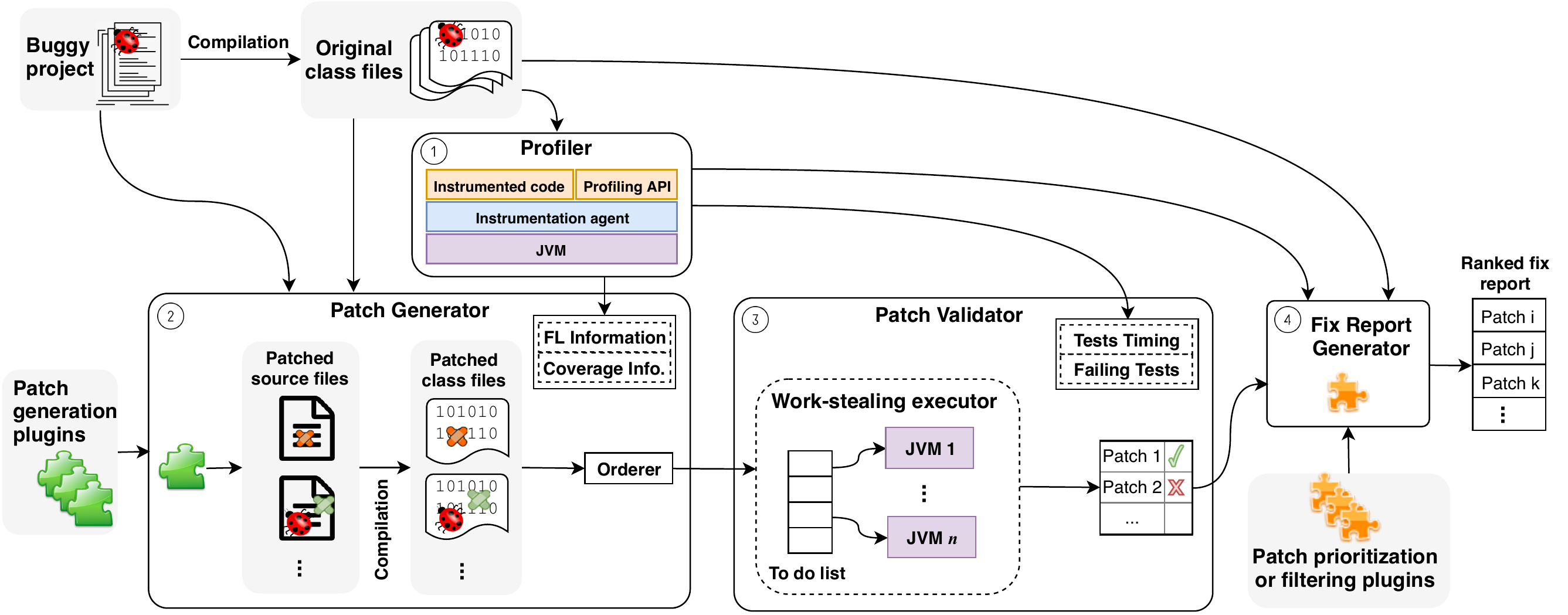}
    \caption{An overview of \prf}
    \label{fig:architecture}
    \vspace{-0.4cm}
\end{figure*}
\input{sections/introduction}

\input{sections/approach}

\input{sections/demo}

\input{sections/related}

\input{sections/conclusions}

\bibliographystyle{ACM-Reference-Format}
\bibliography{bibdb}
\end{document}

%% file: misc/commands.tex
\newcommand{\gv}{G\&V\xspace}

\newcommand{\prf}{PRF\xspace}
\newcommand{\capgen}{CapGen\xspace}
\newcommand{\elixir}{ELIXIR\xspace}
\newcommand{\flair}{FLAIR\xspace}
\newcommand{\astor}{ASTOR\xspace}
\newcommand{\objsim}{ObjSim\xspace}
\newcommand{\prapr}{PraPR\xspace}
\newcommand{\junit}{JUnit\xspace}

\crefformat{section}{\S#2#1#3}
\crefformat{subsection}{\S#2#1#3}
\crefformat{subsubsection}{\S#2#1#3}

\definecolor{darkblue}{rgb}{0.0,0.0,0.6}
\lstdefinelanguage{XML}
{
  morestring=[b]",
  morestring=[s]{>}{<},
  morecomment=[s]{<?}{?>},
  stringstyle=\color{black},
  identifierstyle=\color{cyan},
  keywordstyle=\color{darkblue},
  showstringspaces=false,
  basicstyle={\small\ttfamily},
  morekeywords={artifactId,version,groupId,plugin,configuration,failingTests}%
}

\lstdefinelanguage{CSV}
{
  basicstyle={\small\ttfamily},
}

\newcommand{\Comment}[1]{}

\newcommand{\ali}[1]{\textcolor[rgb]{0.0,0.0,1.0}{#1}}
\newcommand{\andi}[1]{\textcolor[rgb]{1.0,0.0,0.0}{#1}}

%% file: sections/introduction.tex
\vspace{-0.1cm}
\section{Introduction}\label{sec:intro}
Automatic program repair (APR) \cite{apr2019cacm} is one of the recent advances in automated software engineering that holds the promise of reducing debugging costs by suggesting high-quality patches that either directly fix the bugs or help the developers during manual debugging \cite{liang2020interactive}.
In the last decade, APR has been the subject of intense research and still remains an active area of research \cite{bib:GMM17,monperrus2018living}.

Generate-and-validate (\gv) refers to the class of APR techniques that attempt to fix the bug by repeatedly generating patches to produce program variants that subsequently get validated against certain rules or checks.
A patch is called \emph{plausible} if it passes all the checks.
Validating a patch can be accomplished via a spectrum of techniques ranging from sound formal verification techniques \cite{van2018static} to testing \cite{astor16,ghanbari2019,wen18}. 
However, in real-world situations, formal specification of software is usually absent and automating formal verification is theoretically impossible.
Testing, on the other hand, remains as the prevalent, economic method of getting more confidence about the quality of software, and a vast majority of \gv APR techniques use tests as correctness criteria.

By examining the architecture of state-of-the-art test-based \gv APR tools \cite{bib:GMM17}, we observed that most of them have components for fault localization (FL) \cite{wong2016survey}, patch generation, and patch filtering/prioritization. 
During fault localization a \textit{suspiciousness value} is assigned to each executed program location.
During patch generation, a subset of these locations are transformed, resulting in a set of patches.
Since test cases do not perfectly specify the software, many of the generated patches happen to pass all the tests without fixing the bugs.
Such patches are known as \textit{overfitted} patches \cite{overfitting15}.
Therefore, the APR process is usually followed by a post-processing phase to filter out overfitted patches or prioritize patches that are more likely to be correct.
Different program repair techniques usually differ in the way they generate patches to produce program variants, while the other components remain more or less the same and are reused from one implementation to another or reinvented.
Even reusing these processes takes a considerable amount of programming effort for the APR researchers when building their prototypes.

This paper presents \prf (\textbf{P}rogram \textbf{R}epair \textbf{F}ramework), a framework for building APR prototypes by extending it with a patch generation plugin for transforming the identified suspicious locations.
\prf provides the APR tool programmer with spectrum-based FL information at different levels of granularity (\textit{e.g.,} line-/statement-, method-, or class-level). 
The framework also provides a safe, fast, and configurable facility for patch validation, and a customizable fix report generation component for prioritization/filtering of plausible patches. 
The patch validation component employs an assortment of techniques to safely accelerate patch validation process; it reorders test cases and does test selection.
The component also implements a novel work-stealing algorithm \cite{wiki:workstealing} so that the patches can be validated concurrently, thereby maximizing CPU utilization. 
This compensates for the overhead of repeated creation of Java Virtual Machine (JVM) processes to achieve safety in patch validation.
Although \prf is shipped with a default fix report generation plugin, the APR tool programmers can construct their own patch prioritization/filtering plugins for a customized fix report generation.

\prf depends solely on runtime information, so it is useful not just for prototyping Java APR tools but also for building tools targeting other JVM-based programming languages such as Kotlin, Scala, etc. 
It also comes in the form of a Maven plugin so that it can be reliably applied on different Maven-based projects.

We have constructed a patch generation plugin by specializing \capgen APR tool proposed by Wen \textit{et al.} \cite{wen18}.
Our experiments with \capgen show that patch validation is responsible for 77.7\% of the total repair time, on average, and our integration with \prf results in 3.1X speedup when reordering test cases while validating patches sequentially without any test selection. 
In addition, we obtained an 11.4X speedup in patch validation when using 8 CPU cores combined only with test reordering. 

Constructing a plugin for \capgen is as simple as making a wrapper for the tool and disabling the patch validation and prioritization processes that are to be delegated to \prf.
We expect that with the help of APR libraries like \astor \cite{astor16}, APR researchers will be able to make fast and reliable prototypes easily. 
This will, in turn, greatly help the reproducibility of APR experiments. 
Such a customizable framework has the added benefit of simplifying studies that compare different algorithms used in certain aspects of APR (\textit{e.g.}, fault localization \cite{liu2019you} or patch classification/prioritization \cite{ye2019automatedb}).

\prf and its documentation, together with experimental result for \capgen, are publicly available on GitHub \cite{prf}.

\Comment{ \ali{I think these could be useful for advertising our work. I am submitting an initial version, later let me know if you want me to move the text as described by you. We can update the submission.}\andi{I would move the last two paragraphs into the next section. Maybe keep the first of the two paragraphs, but substantially reduced. All the details should be in the next section.}}

%% file: sections/approach.tex
\vspace{-0.2cm}
\section{Overview of \prf}\label{sec:overview}
Inspired by the steps common to state-of-the-art test-based \gv APR techniques, we build \prf with four components responsible for profiling, patch generation, patch validation, and fix report generation. 
Figure \ref{fig:architecture} shows \prf's main components.
We next describe each component in more details.

\vspace{-0.2cm}
\subsection{Profiler}
Any non-trivial \gv APR tool needs information about the program under repair so as to make sensible transformations, validate the generated patches efficiently, or to generate high-quality fix reports.
For example, in most cases, it makes sense to mutate only suspicious program locations, and in order to achieve performance in patch validation the tool might reorder test cases based on their execution time.
We use the term \textit{profiling} to refer to the task of obtaining information about the program under repair.
These include test results and execution time, coverage information, dynamic call graph, and fault localization information that are used in different components for various purposes such as patch generation, test selection, test reordering, and patch prioritization.

Since APR tool programmer, or the end users, might opt for other sources for these information, except for test execution time measurement and recording test results, which becomes freely available once \prf executes the tests, all features of the \textit{profiler} component are optional.

During profiling, the program is instrumented for recording the aforementioned runtime information. 
\prf uses Java Agent technology and the ASM bytecode manipulation framework \cite{objectweb:asm} to instrument the input program. 
Besides constructing dynamic call graph and recording test results, test execution time and, depending on the user configuration, coverage information of test cases at different levels of granularity (\textit{i.e.}, class-, method-, or line-level) are recorded. 
If the user selects a certain level of granularity and specifies a certain FL formula \cite{wong2016survey}, then the collected coverage information is used to calculate the suspiciousness values for the program elements based on the specified formula.

\vspace{-1.1cm}
\subsection{Patch Generator}
Patch generation is an integral part of \gv APR algorithms wherein the patches are constructed via transforming a subset of program locations.
Different APR algorithms mostly differ in the way they generate the patches, and the philosophy behind \prf is to separate patch generation phase from the rest of phases and create a generic, customizable program repair framework.

In \prf, the \emph{patch generator} component is responsible for generating patches which is customizable via a user provided \textit{patch generation plugin}. 
Depending on the patch generation algorithm, the program source code and/or compiled class files of the original buggy program are transformed by the specified patch generation plugin. 
Thus, the framework passes the path name of the source files, test sources, and compiled binaries to the plugin.
Additionally, depending on which feature of the profiler component are activated, test coverage information, dynamic call graph, and FL information with the specified level of granularity shall be fed to the plugin.

The patch generation plugin is intended to generate a pool of patches that are stored on the disk.
\prf expects the patches to be compiled into class files and it is the responsibility of the patch generation plugin to compile the generated patches using the appropriate JVM-based compiler. 
Furthermore, in order to do test selection in patch validation phase, \prf relies on the patch generation plugin to determine which test cases cover the patched location for each patch. 
If the plugin does not specify the covering tests for a given patch, then during the subsequent patch validation phase, all the test cases shall be executed against the patch and no test selection will take place. 

\vspace{-0.1cm}
\subsection{Patch Validator}\label{sec:patchval}
The patch generator component sends the list of generated patches to the \emph{patch validator} component wherein (a subset of) the test cases are executed against the patches to identify plausible patches. 
Depending on the degree of parallelism specified by the user, the component validates the patches sequentially or in parallel. 
Furthermore, depending on the user preferences, patch validation can be stopped after the first plausible patch is found, or the entire search space will be explored and the identified plausible patches will get prioritized and/or filtered later.

\begin{figure}
    \centering
    \includegraphics[scale=0.38]{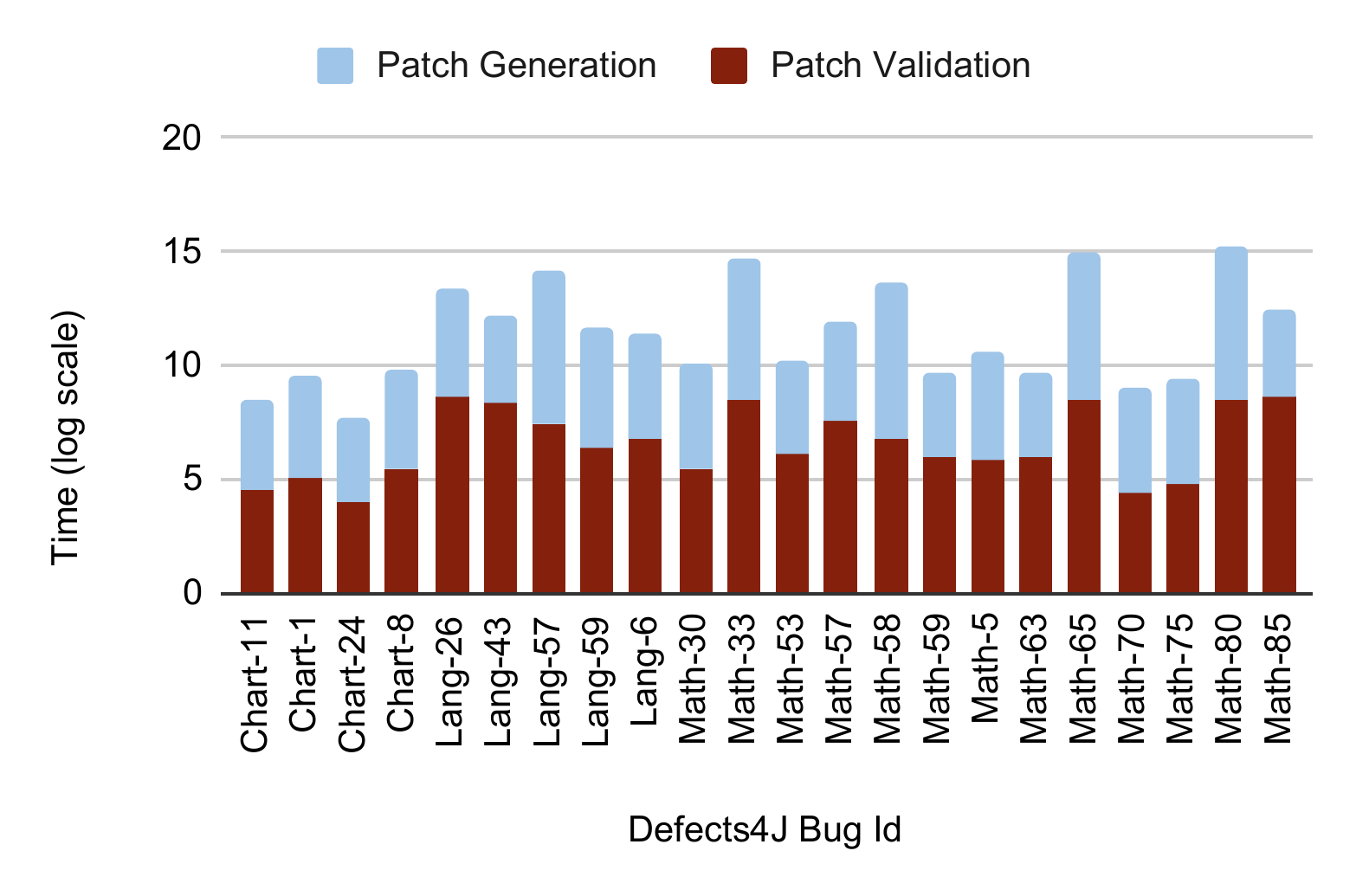}
    \caption{Patch generation time for \capgen and validation time for the generated patches, for 22 bugs in Defect4J.}
    \label{fig:chartTime}
    \vspace{-0.4cm}
\end{figure}

It is worth noting that patch validation takes up a large portion of the end-to-end repair time.
For example, Figure \ref{fig:chartTime} shows the time that an existing APR tool, \capgen \cite{wen18}, spends on generating patches vs. on validating the generated patches, for each of the 22 Defects4J \cite{just2014defects4j} bugs that it is able to fix. 
Our measurements indicate that patch validation takes up 77.7\% of total repair time, on average. 
Similar observations are also reported in previous research by Mehne et al. \cite{mehne2018accelerating}. 
As we shall discuss in \cref{sec:related}, there are a number of methods to reduce patch validation time, but in this work we employ a novel approach: (1) each patch is validated in a separate process; (2) only the tests covering patched location are executed; (3) unless otherwise instructed by the user, \prf reorders test cases to run shorter tests first and also originally failing tests before originally passing ones; (4) unless otherwise instructed by the user, \prf runs patch validating processes in parallel.

Since patch validation involves running test cases and test execution have side-effects, \prf validates each patch in a separate JVM instance to contain the side-effects as much as possible.
Repeated creation of JVM instances is expensive as initializing a JVM session, besides process creation overhead, involves costly tasks of class loading, linking, and JIT-optimization.
These costs are compensated for by the speedup gained through test selection, test reordering, and parallelism.

By using test selection and test reordering methods from regression testing, which have been successfully applied in APR \cite{ghanbari2019,mehne2018accelerating}, \prf runs only the test cases that cover the patched locations and runs the failing test cases before the passing ones as they are more likely to fail again and invalidate the incorrect patch as soon as possible.
\prf aims for further speeding up patch validation process by overlapping JVM process creation with the effective work done during patch validation phase, namely running test cases.
This is achieved via an implementation of work-stealing algorithm \cite{wiki:workstealing} which ensures CPU utilization is maximized by always keeping its cores busy doing different tasks in parallel.

Sometimes patching may create infinite loops and this causes tests to run forever during patch validation.
To avoid this issue, \prf uses test execution time recorded during profiling phase to identify patches that are likely to have entered an infinite loop, by setting a timeout for each test case. 
We use the formula $\beta+(1+\alpha)\tau_t$ to calculate the time budget for a test $t$, based on the original execution time $\tau_t$ of the test and the user-defined parameters $\alpha$ and $\beta$. 
Following heuristics determined during our experiments, we set $\beta=5,000$ and $\alpha=0.5$, meaning that a test taking more than 1.5 times its original execution time plus 5 seconds is deemed timed out. 
A patch for which at least one test times out, is deemed non-plausible and will not be passed to the next component.

\subsection{Fix Report Generator}
The \textit{fix report generator} component produces a list of patches to be examined by the user of APR techniques.
It may use a patch prioritization and/or filtering plugin to first display the patches that are more likely to be correct or filter out plausible but likely incorrect patches. 
Such a plugin would be provided by the user.
The current implementation of the fix report generator prints the list of plausible patches in an arbitrary order which might be time consuming and boring.
We plan to integrate our JVM language agnostic patch prioritization tool \objsim \cite{ghanbari20objsim}, which has shown promising results in our experiments with the APR tool \prapr \cite{ghanbari2019}.

%% file: sections/demo.tex
\input{tables/options}
\vspace{-0.1cm}
\section{\prf Usage}\label{sec:usage}
\prf comes in the form of a Maven plugin. 
After checking out the source code from the GitHub repository \cite{prf} and installing the Maven plugin and the core library for \prf on the local Maven repository, the framework will be ready to use. 
The following XML snippet shows the minimal amount of configuration in the POM file of the target buggy project. 
\vspace{-0.1cm}
\begin{lstlisting}[language=xml]
<plugin>
    <artifactId>prf-maven-plugin</artifactId>
    <groupId>edu.utdallas</groupId>
    <version>1.0-SNAPSHOT</version>
</plugin>
\end{lstlisting}
\vspace{-0.1cm}

Once \texttt{mvn edu.utdallas:prf-maven-plugin:run} is executed through the command-line, \prf shall use a default patch generation plugin named \texttt{DummyPatchGenerationPlugin} which simply looks for a directory named \texttt{patches-pool} under the base directory of the project. 
This directory is expected to contain the pool of the generated patches, and the class file(s) for each patch must reside in a separate sub-directory. 
Names of the sub-directories shall be used as patch identifiers during the fix report generation.
Similarly, \prf uses a default patch prioritization plugin named \texttt{DummyPatchPrioritizationPlugin} that does not have any effect.
Furthermore, by default, only test execution time and test result collector is active in the profiler, and patch validator uses all the available CPU cores to validate patches in parallel.

These default choices can be overridden by the user through the POM file for the program under repair.
Such configurations go under the \texttt{<configuration>} tag in plugin description in the above XML code.
Table \ref{tab:options} summarizes the options that the users can tune.
More details about using \prf is available in the project documentation \cite{prf} and in the companion demo video \url{https://bit.ly/3ehduSS}.

%% file: tables/options.tex
\begin{table*}[t!]
    \centering
    \caption{Summary of \prf options configurable through the POM file}
    \label{tab:options}
    \begin{adjustbox}{width=0.8\textwidth}
    \begin{tabular}{c||l}
        \textbf{Option} & \multicolumn{1}{c}{\textbf{Descriptions}} \\\hline\hline
        \texttt{<flOptions>} & Takes values \texttt{OFF} for no FL, or \texttt{CLASS\_LEVEL}, \texttt{METHOD\_LEVEL}, or \texttt{LINE\_LEVEL} for class-/method-/line-level FL\\
        \texttt{<flStrategy>} & Takes the values \texttt{OCHIAI} or \texttt{TARANTULA}, for Ochiai or Tarantula spectrum-based FL \cite{wong2016survey}, respectively \\
        \texttt{<testCoverage>} & If true, line-level coverage information for tests will be collected and passed to other components \\
        \texttt{<failingTests>} & If left empty, failing tests will be inferred. Each failing should be in a \texttt{<failingTest>} tag \\
        \hline
        \texttt{<cgOptions>} & Takes values \texttt{OFF} (default) or \texttt{DYNAMIC} to enable/deactivate dynamic call graph construction \\
        \hline
        \texttt{<patchGenerationPlugin>} & The name of patch generation plugin. By default, this is set to \texttt{dummy-patch-generation-plugin} \\
        \texttt{<parallelism>} & Degree of parallelism for patch validation. By default, it is 0, and all CPU cores will be used \\
        \hline
        \texttt{<patchPrioritizationPlugin>} & The name of patch prioritization plugin. By default, this is set to \texttt{dummy-patch-prioritization-plugin} \\ \hline
        \texttt{<timeoutConstant>} & Constant part of the timeout value calculation, \textit{i.e.,} $\beta$ in \cref{sec:patchval} \\
        \texttt{<timeoutPercent>} & Percent part of the timeout value calculation, \textit{i.e.,} $\alpha$ in \cref{sec:patchval}
    \end{tabular}
    \end{adjustbox}
\end{table*}

%% file: sections/related.tex
\section{Related Work}\label{sec:related}
\flair is a proprietary framework used in Fujitsu Labs America to construct the APR system \elixir \cite{saha2017elixir}.
However, this system is not publicly available.
\prf is the first open-source framework that provides all the functionalities offered by a framework like \flair, plus \prf offers an efficient patch validation facility.

\astor \cite{astor16} is a general-purpose library for developing source code level, Java-based APR tools. 
Unlike \astor, \prf is a customizable framework upon which different patch generation algorithms can be installed as plugins and different strategies for patch prioritization and/or filtering can be employed. 
\prf is not intended to replace \astor, instead, \prf complements the library in that it enables APR researchers focus on taking full advantage of \astor's features to implement more reliable patch generation algorithms, and better understand the impact of different fault localization, patch prioritization and/or filtering strategies on the effectiveness of their implementation.
As a toolset for prototyping APR techniques, \prf can be seen as a step similar in nature to \cite{d4jChallenge}.

Le Goues \textit{et al.} \cite{le2013current} highlight the high cost of patch validation in test based \gv APR. 
Mehne et al. \cite{mehne2018accelerating} report that patch validation can take between 40\% to 92\% of total repair time and propose to prune the patches needed to be tested as well as test case selection to reduce this cost.
A recent line of research \cite{ghanbari2019,guo2019speedup}, proposes to use the HotSwap trick offered by the JVM to validate the patches on-the-fly, without restarting the JVM.
This was previously used in mutation testing systems like PIT \cite{coles2016pit}.
It has the benefit of avoiding the high costs of restarting JVM for each patch, but the patches that involve altering the class structure (\textit{e.g.,} addition or removal of class members) are not eligible for being HotSwapped.
The approach in order to be effective needs proper isolation of test execution side-effects (\emph{e.g.,} \cite{bell2014unit}).
However, such a method does not solve the current limitations of a HotSwap-based approach.
In this work, we follow a different approach that proves to be fast, reliable, and more effective.
\prf validates each patch in a fresh JVM session, thereby containing the side-effects of test execution and also avoiding restrictions of a HotSwap-based approach, and instead of dealing with the internals of JVM, we rely on test selection and reordering, as well as parallelism to speedup patch validation.

%% file: sections/conclusions.tex
\vspace{-0.2cm}
\section{Conclusions and Future Work}\label{sec:conclusions}
With \prf, APR researchers will be able to build research prototypes by simply providing a patch generation plugin and fine tuning the framework's existing components for multi-granularity level fault localization, patch validation, and fix report generation.
\prf uses a novel patch validation technique, relying on test selection and prioritization as well as parallelism, that achieves 11+X speedup compared to vanilla testing.
\prf is publicly available at \cite{prf}.

We are working on integrating our JVM language agnostic patch prioritization system \objsim \cite{ghanbari20objsim} in \prf and release the framework with an effective built-in patch prioritization mechanism.